\documentclass[conference]{IEEEtran}
\IEEEoverridecommandlockouts
\usepackage{cite}
\usepackage{amsmath,amssymb,amsfonts}
\usepackage{graphicx,subcaption}
\usepackage{textcomp}
\usepackage{xcolor}
\usepackage[ruled,linesnumbered]{algorithm2e}
\usepackage[]{hyperref}
\usepackage[T1]{fontenc}
\usepackage{bm}

\def\RR{\mathbb{R}}

\def\CC{\mathbb{C}}

\def\SSH{\mbox{S-SPAIN\,H}}
\def\SSOMP{\mbox{S-SPAIN\,OMP}}

\DeclareTextAccent{\ring}{OT1}{23}

\def\x{\vect{x}}
\def\u{\vect{u}}

\def\y{\vect{y}}
\def\z{\vect{z}}

\def\snr{SNR}

\newcommand{\norm}[1]{\|#1\|}

\newcommand{\adjoint}[1]{#1^*}

\newcommand{\vect}[1]{\mathbf{#1}} 
\newcommand{\matr}[1]{\mathbf{#1}} 

\newcommand{\argmin}{\mathop{\operatorname{arg~min}}}
\newcommand{\modulo}{\mathop{\operatorname{mod}}}

\newcommand{\qm}[1]{``#1''}  

\newcommand{\ig}{\iota_{\Gamma}(\x)}
\newcommand{\il}{\iota_{\ell_0 \leq k}(\z)}
\def\BibTeX{{\rm B\kern-.05em{\sc i\kern-.025em b}\kern-.08em
    T\kern-.1667em\lower.7ex\hbox{E}\kern-.125emX}}

\begin{document}

\title{Introducing SPAIN (SParse Audio INpainter)\\
\thanks{
	The work was supported by the joint project of the FWF and the Czech Science Foundation (GA\v{C}R): numbers I 3067-N30 and 17-33798L, respectively. 
	Research described in this paper was financed by the National Sustainability Program under grant LO1401. 
	Infrastructure of the SIX Center was used.}
}


\author{\textit{Ond\v{r}ej Mokr\'y}$^{\star,\dagger}$ \qquad 
	\textit{Pavel Z\'avi\v{s}ka}$^{\star}$ \qquad 
	\textit{Pavel Rajmic}$^{\star}$ \qquad 	
	\textit{V\'it\v{e}zslav Vesel\'y}$^{\dagger}$
	
	\\ \\
	
	$^{\star}$ Signal Processing Laboratory, Brno University of Technology, Brno, Czech Republic \\
	$^{\dagger}$ Faculty of Mechanical Engineering, Brno University of Technology, Brno, Czech Republic \\
	Email: \{170583, xzavis01, rajmic, vitezslav.vesely\}@vutbr.cz \\
}

\maketitle

\begin{abstract}
A novel sparsity-based algorithm for audio inpainting is proposed.
It is an adaptation of the SPADE algorithm by Kiti\'{c} et\ al., originally developed for audio declipping, to the task of audio inpainting.
The new SPAIN (SParse Audio INpainter) comes in synthesis and analysis variants.
Experiments show that both \mbox{A-SPAIN} and S-SPAIN outperform other sparsity-based inpainting algorithms.
Moreover, A-SPAIN performs on a par with the state-of-the-art method based on linear prediction in terms of the SNR,
and, for larger gaps, SPAIN is even slightly better in terms of the PEMO-Q psychoacoustic criterion.
\end{abstract}


\section{Introduction}
The term \qm{inpainting} has spread to the audio processing community from the image processing field, when a paper by Adler et\ al.\ was published, entitled simply \qm{Audio inpainting} \cite{Adler2012:Audio.inpainting}.
The goal of so termed restoration task is to fill degraded or missing parts of data, based on its reliable part, and preferably in an~unobtrusive way.
Although the term \qm{inpainting} is rather new in this context, the problem itself is much older and different approaches have been proposed to deal with it in past decades.


One of the first successful approaches was the time-domain interpolation of the missing audio samples.
An adaptive interpolation method was presented by Janssen et\ al.\ in \cite{javevr86}, which was built on modeling the signal as an~autoregressive (AR) process.
The method can be roughly described as the minimization of a~suitable functional, which includes as the variables both the estimates of the missing samples and the AR coefficients of the restored signal.
Although the algorithm is designed as iterative, only a~single iteration was considered in \cite{javevr86}.
On today's hardware, nonetheless, hundreds of iterations can be computed in a reasonably short time.
Doing this improves the results substantially and keeps the Janssen algorithm the state-of-the-art in audio inpainting in terms of the \snr.
Note, however, that when applied to compact gaps, the AR-modeling suggests that this method is successful for sounds that contain harmonics which do not evolve within the gap.
As such, this approach is beneficial for gaps of up to ca 45 milliseconds, which can be observed also from the experiments described below.
Various prediction methods were also presented in \cite{fink2013:extrapolation}.

A different approach is the sparsity-based audio inpainting, reported more recently in \cite{Adler2012:Audio.inpainting}.
It benefits from the observation that the energy of an audio signal is often concentrated in a~relatively small number of coefficients, with respect to a~suitable, redundant time-frequency transform.
The goal is to find a~signal that has sparse representation under a~transform such as the STFT (Short-Time Fourier Transform, aka the Discrete Gabor Transform, DGT),
while it belongs to the set of feasible solutions;
for audio inpainting, a~solution is feasible if it is identical to the reliable parts of the signal.

The above requirements can be formulated as an 
optimization task.
Its solution cannot be attained in practice due to the NP-hardness, but the solution can be reasonably approximated, 
for example using the Orthogonal Matching Pursuit (OMP) as in \cite{Adler2012:Audio.inpainting}.
Another kind of approximation uses the $\ell_1$ norm, which leads to a~convenient convex minimization
\cite{Lieb2018:Audio.Inpainting}. 
One of the modern approaches is also to search for sparsity in a~non-local way \cite{ToumiEmiya2018:Sparse.Non.Local.Inpainting}, or using statistical prior information about the sparse representation \cite{Adler2016:CAMP}.

Further methods have been developed to deal with long missing segments of audio signal.
Stationary signals can be restored using sinusoidal modeling \cite{lagrange2005long} or via model-based interpolation schemes \cite{Esquef2003:Interpolation.Long.Gaps.Warped.Burgs}.
In repetitive signals such as rock and pop music,
there is usually an intact part sufficiently similar to the distorted part;
this fact is utilized for filling the missing data in \cite{Perraudin2018:Similarity.Graphs}.
A~different method, still based on self-similarity, was introduced in \cite{Bahat_2015:Self.content.based.audio.inpaint}, aiming originally at the packet loss concealment. 
One of the latest methods employs a deep neural network for the task of audio inpainting
\cite{MarafiotiPerraudinHolighausMajdak2018:Context.encoder.inpainting}.


The algorithm presented in the paper is inspired by the state-of-the-art among sparsity-based approaches to audio declipping, the so-called SParse Audio DEclipper (SPADE)
\cite{Kitic2015:Sparsity.cosparsity.declipping,ZaviskaRajmicPrusaVesely2018:RevisitingSSPADE}
and it is in line with the universal framework presented in \cite{GaultierBertinKiticGribonval2017:Framework.audio.restoration}.
We benefit from the close relation between the sparsity-based inpainting and declipping problems. 
%
We formulate the two variants of the inpainting problem as the optimization tasks
\begin{subequations}
	\label{eq:problem_SPAIN}
	\begin{align}
	\label{subeq:analysis}
	&\min_{\x,\z} \norm{\z}_0\quad\text{s.\,t.}\quad\x\in\Gamma\ \,\text{and} \ \,\norm{A\x-\z}_2\leq\varepsilon, \\
	\label{subeq:synthesis}
	&\min_{\x,\z} \norm{\z}_0\quad\text{s.\,t.}\quad\x\in\Gamma\ \,\text{and} \ \,\norm{\x-D\z}_2\leq\varepsilon,
	\end{align}
\end{subequations}
where \eqref{subeq:analysis} and \eqref{subeq:synthesis} present the formulation referred to as the analysis and the synthesis variant, respectively.
In both formulations, $\Gamma = \Gamma(\y)\subset \CC^N$ is the set of feasible solutions
(i.e.\ the set of time-domain signals that are equal to the observed signal $\y$ in its reliable parts),
$A:\CC^N\to\CC^P$ is the analysis operator of a Parseval tight frame and $D = \adjoint{A}$ is its synthesis counterpart,
with $P\geq N$
\cite{christensen2008}.
The formulation \eqref{eq:problem_SPAIN} reflects the goal to find a~signal from $\Gamma$ with the highest-sparsity representation, as was informally stated above.

The intention is to make use of the Alternating Direction Method of Multipliers (ADMM) for the optimization of the above non-convex problems, as was the case in the original SPADE paper
\cite{Kitic2015:Sparsity.cosparsity.declipping}.
For this purpose, a~more appropriate formulation is needed.
We introduce the (unknown) sparsity $k$, which is to be minimized, and indicator functions\footnote{The indicator function of a set $\Omega$, denoted $\iota_\Omega(\x)$, attains the value 0 for $\x\in \Omega$ and $\infty$ otherwise \cite{combettes2011proximal}.} of two sets:
the set of feasible solutions $\Gamma$ and the set of $k$-sparse vectors, the latter denoted $\ell_0\leq k$ for short.
This leads to	
\begin{equation}
\min_{\x, \z, k}\{ \ig + \il \} \hspace{1em}\text{s.t.}\hspace{1em}
\begin{cases}
A\x - \z \hspace{-0.5em}&= \mathbf{0},\\
\x - D\z \hspace{-0.5em}&= \mathbf{0}.
\end{cases}
\label{eq:problem_SPAIN_ADMM}
\end{equation}
Note that in \eqref{eq:problem_SPAIN_ADMM}, a~more conservative constraint binding the variables $\x$ and $\z$ is used compared to \eqref{eq:problem_SPAIN}.
The reason is that such a~constraint falls within the general ADMM scheme \cite{Boyd2011ADMM}.
Such an alteration is nevertheless legitimate, since in the resulting iterative algorithm presented below,
the stopping criterion will in effect relax the constraint into the form of \eqref{eq:problem_SPAIN}.


\section{SPAIN (Sparse Audio Inpainter)}
\label{sec:SPAIN}

In \cite{Kitic2015:Sparsity.cosparsity.declipping}, the SPADE algorithm was introduced to tackle \eqref{eq:problem_SPAIN_ADMM}.
It is built on the idea that for a fixed $k$, problem \eqref{eq:problem_SPAIN_ADMM} can be approximately solved by
the ADMM.
The SPADE increases the value of $k$ during the iterations of ADMM (starting from a~sufficiently small value), until the condition $\norm{A\x-\z}_2\leq\varepsilon$ (analysis approach, A-SPADE) or $\norm{\x-D\z}_2\leq\varepsilon$ (synthesis approach, S-SPADE) is met for the chosen tolerance $\varepsilon>0$.
This ensures $k$-sparsity of the signal coefficients during iterations and the algorithm thus provides an approximation of the solution to \eqref{eq:problem_SPAIN}.

Section~\ref{subsec:ASPAIN_from_SPADE} shows that A-SPAIN can be derived from \mbox{A-SPADE}. 
In Subsection~\ref{subsec:SSPAIN_from_ADMM}, on the other hand, S-SPAIN is introduced as a~brand new inpainting algorithm.
The derivation is in line with the new variant of S-SPADE for declipping, proposed in
\cite{ZaviskaRajmicMokryPrusa2019:SSPADE_ICASSP}.

\subsection{A-SPAIN derived from A-SPADE}
\label{subsec:ASPAIN_from_SPADE}

The A-SPADE algorithm (Alg.\,\ref{alg:aspain}) was originally designed for the declipping task,
where the set of feasible solutions $\Gamma$ is the (convex) set of time-domain signals whose samples are identical to the observed ones in the reliable part,
while the restored samples are required to lie above or below the upper or the lower clipping thresholds, respectively;
see \cite{Kitic2015:Sparsity.cosparsity.declipping} or \cite{ZaviskaRajmicPrusaVesely2018:RevisitingSSPADE} for more details.

The key observation leading to SPAIN is that the formulations of inpainting and declipping differ 
\emph{only by the definition of the set of feasible solutions} $\Gamma$, which is actually less restrictive for inpainting, since it does not involve any requirements on the samples at unreliable positions.
Alg.\,\ref{alg:aspain} is formulated general enough to cover both the \mbox{A-SPADE} and the new A-SPAIN.
In effect, the actual task being solved is a~matter of the projection on line~3 of the algorithm.


The operator $\mathcal{H}_k$ used in the algorithm denotes hard thresholding, which sets all but $k$ largest elements of the argument to zero. 
Note that in our implementation, the structures of $D$ and $A$ and the fact that a real signal is being processed are taken into account,
leading to thresholding complex-conjugate pairs of coefficients at a~time
instead of thresholding of a~single vector element.


Although the analysis version of the SPADE algorithm was derived from ADMM in \cite{Kitic2015:Sparsity.cosparsity.declipping} based on the formulation \eqref{eq:problem_SPAIN_ADMM}, the synthesis variant was surprisingly built on a~different basis
(see the technical report \cite{ZaviskaMokryRajmic2018:SPADE_DetailedStudy} for further explanation).
A~more consistent approach is to derive the synthesis variant in analogy to the analysis variant.
Thus, in the following subsection, the novel derivation of S-SPAIN from ADMM is briefly described.

\subsection{S-SPAIN derived from ADMM}
\label{subsec:SSPAIN_from_ADMM}


The ADMM is a scheme to solve optimization problems of the form
\begin{equation}
\min_\z f(\z) + g(D\z),
\label{eq:ADMM_problem_formulation_1}
\end{equation}
where
\(D\) is a~linear operator.
The functions $f,g$ are assumed to be real and convex.
Problem \eqref{eq:ADMM_problem_formulation_1} can be reformulated by introducing a~slack variable $\x$
such that $\x=D\z$, leading to
\begin{equation}
\min_{\z,\x} f(\z) + g(\x) \hspace{1em} \text{s.t.} \hspace{1em} \x-D\z=\mathbf{0},
\label{eq:ADMM_problem_formulation_2}
\end{equation}
which corresponds to the synthesis variant of \eqref{eq:problem_SPAIN_ADMM}.
Next, the Augmented Lagrangian is formed as
\begin{equation}
L_\rho(\x,\z,\bm{\lambda}) = f(\x) + g(\z) + \bm{\lambda}^{\top}(\x-D\z) + \frac{\rho}{2}\norm{\x - D\z}^2_2,
\label{eq:augmented_lagrangian}
\end{equation}
where \(\rho > 0\) is called the penalty parameter and $\bm{\lambda}\in\RR^N$ is the dual variable.
The general scheme of ADMM then consists of three steps---minimization of $L_\rho$ over $\z$, minimization of $L_\rho$ over $\x$
and the update of the dual variable $\bm{\lambda}$.

For the purpose of audio inpainting, we set $f(\z) = \il$ and $g(\x) = \ig$.
Note that such a function $f$ is not convex, therefore the conditions of ADMM are not met.
Nevertheless, ADMM may converge even in such a case
and the experiments show that it provides reasonable results for audio inpainting.

It is convenient to introduce the so-called scaled dual variable $\u = \bm{\lambda}/\rho$ at this moment.
After this, it is straightforward to arrive at the ADMM steps for S-SPAIN in the following form:
\begin{subequations}\label{eq:ADMM_SSPAIN}
	\begin{align}
	\label{eq:ADMM_SSPAIN_f}
	\z^{(i+1)} & = \argmin_\z\|D\z - \x^{(i)} + \u^{(i)}\|^2_2 \hspace{0.4em} \text{s.t.} \hspace{0.3em} \|\z\|_0 \leq k, \\
	\label{eq:ADMM_SSPAIN_g}
	\x^{(i+1)} & = \argmin_\x\|D\z^{(i+1)} - \x + \u^{(i)}\|^2_2 \hspace{0.5em} \text{s.t.} \hspace{0.5em} \x \in \Gamma,\\
	\label{eq:ADMM_SSPAIN_u}
	\u^{(i+1)} & = \u^{(i)} + D\z^{(i+1)} - \x^{(i+1)}.
	\end{align}
\end{subequations}
In the convex case, the minimizations over $\x$ and over $\z$ can be switched without affecting the convergence of ADMM \cite{Boyd2011ADMM}.
We assume that the convergence will not be violated in the non-convex case either.

The solution to the subproblem \eqref{eq:ADMM_SSPAIN_g} is the projection of $(D\z^{(i+1)} + \u^{(i)})$ onto $\Gamma$,
which is easy to compute in the time domain.	
The subproblem \eqref{eq:ADMM_SSPAIN_f} is more challenging---%
it corresponds to the sparse synthesis approximation, where the goal is to get as close as possible to the signal $(\x^{(i)}-\u^{(i)})$
by synthesis using $D$ in such a way that the expansion coefficients are $k$-sparse.
Such a problem is NP-hard due to the non-orthogonality of $D$ in practice;
therefore approximate solutions are enforced.
One possibility is to utilize hard thresholding and compute
\begin{equation}
\z^{(i+1)} \approx \mathcal{H}_k \! \left(D^*(\x^{(i)} - \u^{(i)})\right).
\label{eq:hard_thresholding}
\end{equation}
In \cite{ZaviskaMokryRajmic2018:SPADE_DetailedStudy} we show that such an approximation is reasonably close to the solution of \eqref{eq:ADMM_SSPAIN_f}.
Note that in A-SPAIN, on the contrary, the thresholding step provides an \emph{exact} solution of the corresponding ADMM subproblem.
Alternatively, $\z^{(i+1)}$ can be approximated using $k$ iterations of OMP \cite{eladbook}, which was not considered neither in the original SPADE \cite{Kitic2015:Sparsity.cosparsity.declipping} nor the novel synthesis version \cite{ZaviskaRajmicMokryPrusa2019:SSPADE_ICASSP}.
Such an approach is, however, computationally much more demanding than the thresholding
(since each iteration of OMP employs one synthesis and one analysis%
)
and the experiments show that although it provides a~better approximation for the solution of \eqref{eq:ADMM_SSPAIN_f}, it surprisingly does not lead to a~better result of the whole \mbox{S-SPAIN} algorithm.

\begin{algorithm}[t]
	\DontPrintSemicolon
	\SetKwInput{KwRequire}{Require}
	\SetKw{KwReturn}{return}	
	\KwRequire{\(A, \y, \Gamma, s, r, \varepsilon\)} \vspace{0.3em}
	\({\hat{\x}^{(0)} = \y, \u^{(0)}=\mathbf{0}, i=0, k=s}\)\;
	\(\bar{\z}^{(i+1)} = \mathcal{H}_k \! \left(A\hat{\x}^{(i)}+\u^{(i)}\right)\)\;
	\({\hat{\x}^{(i+1)} = \argmin_\x{\|A\x-\bar{\z}^{(i+1)}+\u^{(i)}\|_2^2}} \text{\hspace{0.5em}s.t.\hspace{0.5em}}\x \in \Gamma\hspace{-1em}\)\;
	\eIf{\(\|A\hat{\x}^{(i+1)}-\bar{\z}^{(i+1)}\|_2 \leq \varepsilon\)}{\textup{terminate}\;}
	{\(\u^{(i+1)}=\u^{(i)}+A\hat{\x}^{(i+1)}-\bar{\z}^{(i+1)}\)\;
		\(i \leftarrow i+1\)\;
		\If{\(i\modulo r = 0\)}{\(k \leftarrow k+s\)\;}
		go to 2\;
	}
	\KwReturn{\(\hat{\x} = \hat{\x}^{(i+1)}\)}
	\caption{A-SPADE / A-SPAIN, depending on the particular choice of $\Gamma$}
	\label{alg:aspain}
\end{algorithm}

The S-SPAIN algorithm is presented in Alg.\,\ref{alg:sspain}.
In the following section, the S-SPAIN variants using hard thresholding and OMP as an approximation of step 2 of the algorithm will be denoted as \SSH{} and \SSOMP, respectively.

\begin{algorithm}[t]
	\DontPrintSemicolon
	\SetKwInput{KwRequire}{Require}
	\SetKw{KwReturn}{return}		
	\KwRequire{\(D, \y, \Gamma, s, r, \varepsilon\)} \vspace{0.3em}
	\({\hat{\x}^{(0)} = D^*\y, \u^{(0)}=\mathbf{0}, i=0, k=s}\)\;
	\(\bar{\z}^{(i+1)} = \argmin_\z\|D\z - \hat{\x}^{(i)} + \u^{(i)}\|^2_2 \hspace{0.4em} \text{s.t.} \hspace{0.3em} \|\z\|_0 \leq k\)\;
	\({\hat{\x}^{(i+1)} = \argmin_\x{\|D\bar{\z}^{(i+1)} -\x +\u^{(i)}\|_2^2}} \hspace{0.54em}\text{s.t.\hspace{0.5em}}\x \in \Gamma\hspace{-1em}\)\;
	\eIf{\(\|D\bar{\z}^{(i+1)}-\hat{\x}^{(i+1)}\|_2 \leq \varepsilon\)}{\textup{terminate}\;}
	{\(\u^{(i+1)}=\u^{(i)}+D\bar{\z}^{(i+1)}-\hat{\x}^{(i+1)}\)\;
		\(i \leftarrow i+1\)\;
		\If{\(i\modulo r = 0\)}{\(k \leftarrow k+s\)\;}
		go to 2\;
	}
	\KwReturn{\(\hat{\x} = \hat{\x}^{(i+1)}\)}
	\caption{S-SPAIN, task in step 2 is to be approximated either by the hard thresholding or OMP}
	\label{alg:sspain}
\end{algorithm}

\section{Experiments and results}
\label{sec:experiments}

\subsection{Evaluation based on signal-to-noise ratio (SNR)}

For the experiment, ten music recordings sampled at 16\,kHz or 44.1\,kHz were used, covering different degrees of sparsity of the STFT coefficients.
The main source was the signals that were examined in the papers
\cite{Adler2012:Audio.inpainting,Kitic2015:Sparsity.cosparsity.declipping,siedenburg2011:Structured_sparsity}.

In each test instance, the objective was to recover a signal containing six gaps, starting at random points in the signal such that the gaps do not overlap and are not too close to affect the restoration \cite{acceleration_audio_inpainting}.
The gap length was chosen from the~set of 10 available lengths, distributed between 5 and 50\,ms.

As the competitors of SPAIN, we used the Janssen algorithm,	the OMP	and both the synthesis and the analysis approaches of the $\ell_1$ relaxation.\!%
\footnote{%
	Implementation of the Janssen algorithm was taken from the Audio Inpainting Toolbox
	\cite{Adler2012:Audio.inpainting}.
	OMP was implemented using the Sparsify Toolbox
	\cite{SparsifyToolbox}.
	For the synthesis and the analysis $\ell_1$ relaxations, our own implementations of the Douglas-Rachford and the Chambolle-Pock algorithms were used, respectively.}
In OMP and SPAIN, we used the overcomplete DFT with redundancy of the transform set to 2
(meaning that the number of frequency channels is twice the number of signal samples). 
All algorithms
were applied frame-wise: the signal was windowed using the Hann window 64\,ms long with a shift of 16\,ms (i.e.\ 75\% overlap),
and the restored blocks were combined using the overlap-add scheme. 
For the $\ell_1$ relaxation, the DGT was used with the same window parameters and with number of frequency channels corresponding to length of the transform window in samples.
Besides this, weighted $\ell_1$ relaxation was also employed.\!%
\footnote{In weighted $\ell_1$ relaxation, the objective function is $\norm{\matr{W}\cdot}_1$. The diagonal matrix $\matr{W}$ allows us to favor chosen coefficients of the STFT when searching for the restored signal.
The same weighting, i.e.\ according to $\ell_2$ norm of truncated atoms, was proposed in \cite{Adler2012:Audio.inpainting} for OMP.}
As the performance measure, we used the signal-to-noise ratio (\snr),
defined
\begin{equation}
	\mathrm{SNR}\left(\x,\hat{\x}\right) = 10\log_{10}\dfrac{\norm{ \x}_2^2}{\norm{\x-\hat{\x}}_2^2}\quad\text{[dB]},
	\label{eq:SNR}
\end{equation}
where $\hat{\x}$ stands for the~recovered gap and $\x$ denotes the corresponding segment of the~uncorrupted signal
\cite{Adler2012:Audio.inpainting}.

\begin{figure}
	\centering
	\includegraphics[width=0.9\linewidth]{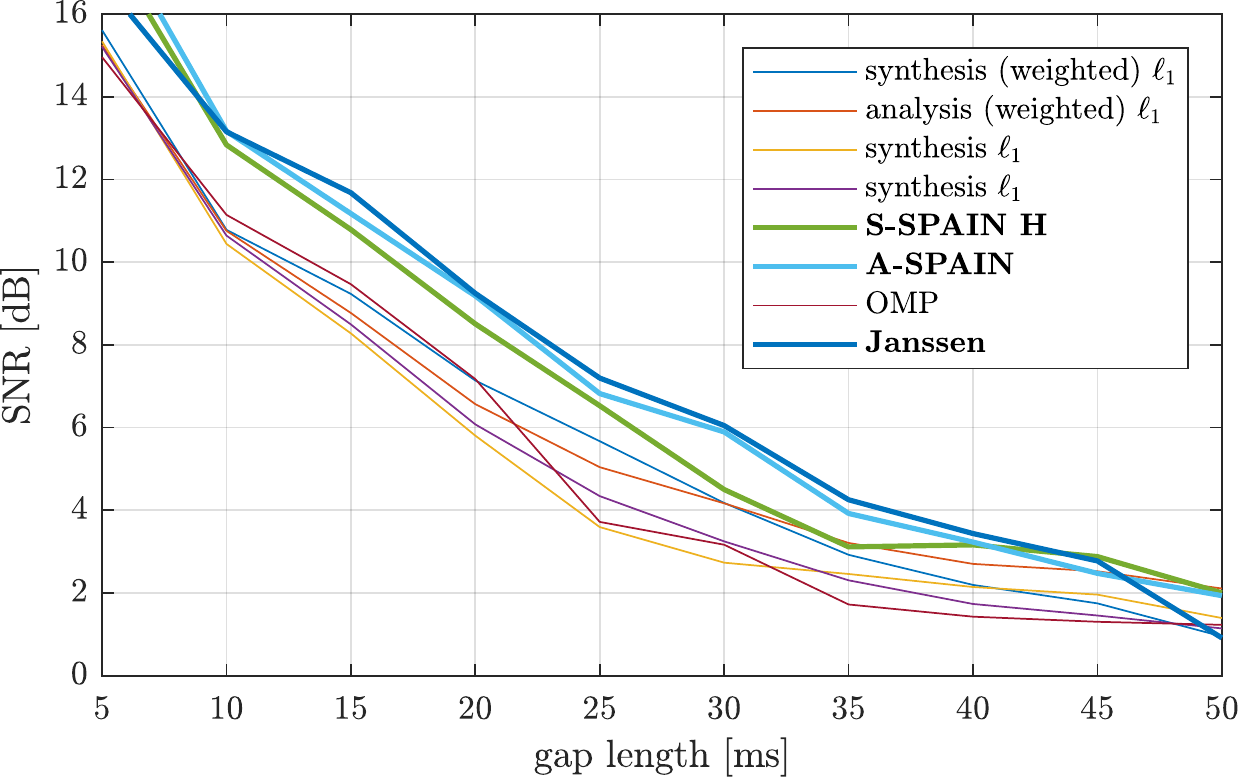}
	\caption{Inpainting results in terms of the SNR.}
	\label{fig:lineplot}
\end{figure}

Fig.~\ref{fig:lineplot} shows the overall results;
for each gap length and each algorithm, the average value of \snr{} was computed from all the signals and all positions of the gap.
It is clear that for gaps of up to 40\,ms, Janssen and A-SPAIN outperform all other methods and that these two algorithms behave almost identically in terms of the SNR
(the biggest difference is for the longest gaps greater than 45\,ms, where the performance of Janssen drops).
Regarding S-SPAIN, the performance is comparable with A-SPAIN and Janssen for shorter gaps (up to 25\,ms), and it outperforms the OMP and the $\ell_1$-relaxation methods even for the longer gaps.
A~possible explanation of the superiority of A-SPAIN over S-SPAIN is that the ADMM subproblems in the analysis variant are solved exactly,
whereas in the synthesis one, the thresholding operator provides just an approximation of \eqref{eq:ADMM_SSPAIN_f}.
Note that in Fig.~\ref{fig:lineplot}, results of \SSOMP{} are not presented.
The reason is the computational complexity of this approach
(a~single test instance takes up to a few days!).

\begin{figure}
	\centering
	\includegraphics[width=0.9\linewidth]{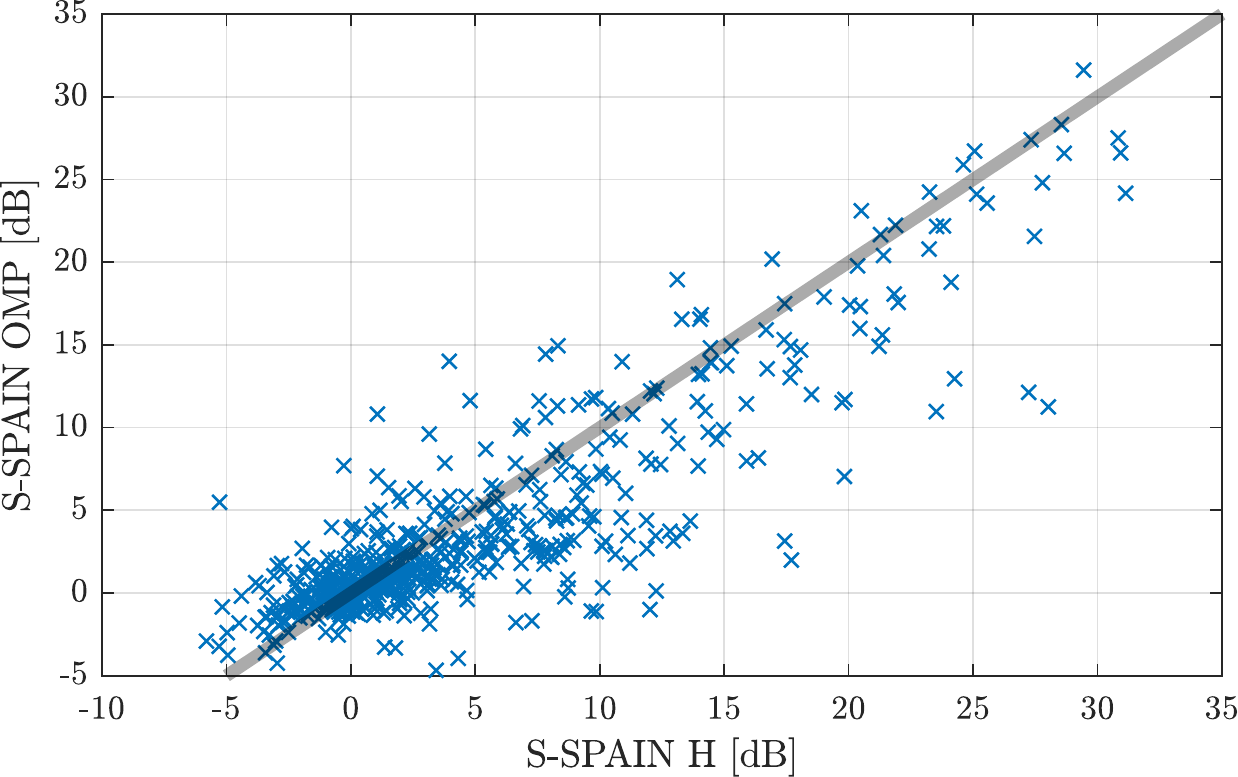}
	\caption{Comparison of \SSH{} and \SSOMP.}
	\label{fig:scatterplot1}
\end{figure}

The scatter plot in Fig.~\ref{fig:scatterplot1} shows detailed comparison of \SSH{} with \SSOMP.
Due to the computational time consumed by the \SSOMP, a~shortened experiment was performed
with both the gap lengths and the window length shortened to their quarter compared to the first experiment
(i.e.\ window length 16\,ms, overlap 4\,ms, gap lengths ranging from 1.25 to 12.5\,ms).
Each cross in Fig.\,\ref{fig:scatterplot1} corresponds to one test instance and its coordinates are \snr{} for \SSH{} and \SSOMP.
The diagonal line in the plot divides the areas in which \SSH{} (under the line) or \SSOMP{} (above the line) perform better.
It is clear that except for a~small area corresponding to a very low \snr, \SSH{} outperforms \SSOMP{} in majority (approx. 58\,\%) of cases.
Also the single-tailed Wilcoxon signed rank test\footnote{Performed using function \texttt{signrank} in MATLAB.}
suggests that the median of results of \SSH{} is greater than in case of \SSOMP{} with significance level 0.05.
This result is quite surprising since the OMP is supposed to solve the optimization subproblem \eqref{eq:ADMM_SSPAIN_f} better than the simpler hard thresholding does.
%

\begin{figure}
	\centering
	\begin{subfigure}{0.45\textwidth}
		\includegraphics[width=\linewidth]{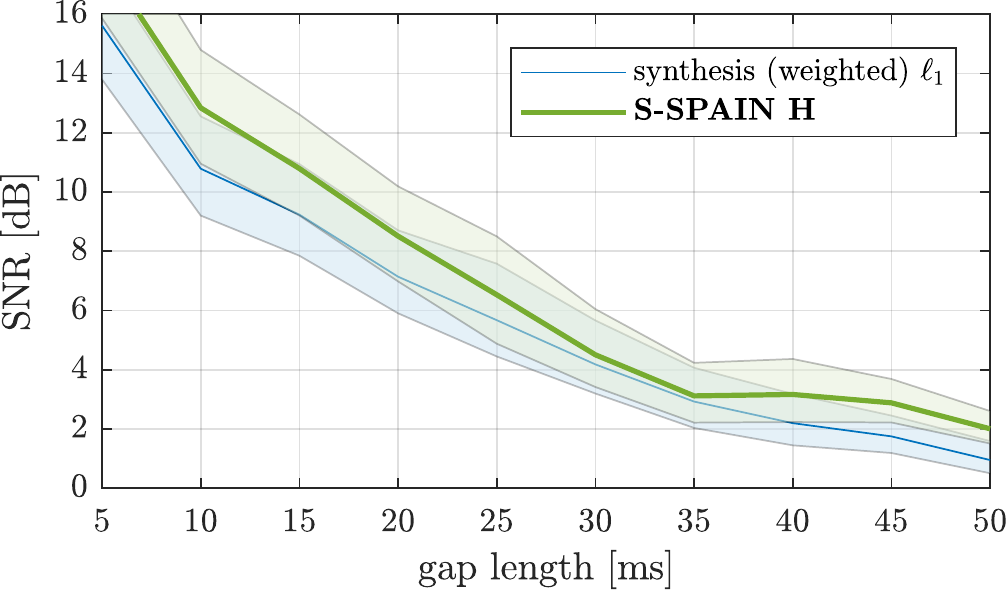}
		\caption{synthesis model}
		\label{fig:bootstrap:synthesis}
	\end{subfigure}\\
	\vspace{0.5em}
	\begin{subfigure}{0.45\textwidth}
		\includegraphics[width=\linewidth]{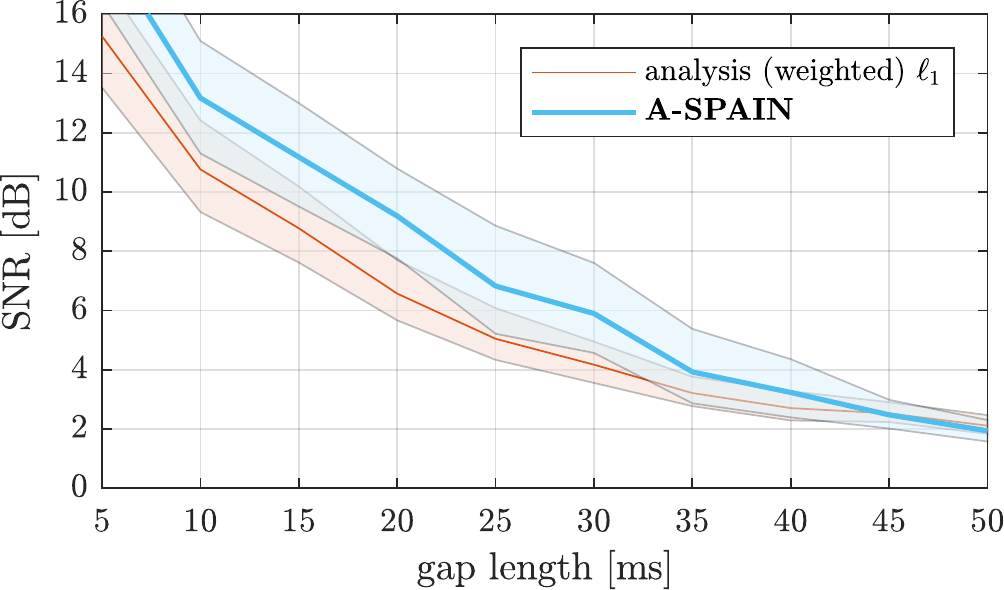}
		\caption{analysis model}
		\label{fig:bootstrap:analysis}
	\end{subfigure}
	\caption{Bootstrap 95\% confidence intervals for the mean value of \snr{} for chosen curves from Fig.\,\ref{fig:lineplot}.
	The estimates with significance level 0.05 was computed using bootstrapping \cite{EfronTibshirani:Bootstrap} with 10\,000 random draws from the population 	for each combination of algorithm and gap length.}
	\vspace{-0.5em}
	\label{fig:bootstrap}
\end{figure}

For a more thorough analysis of the results, Fig.\,\ref{fig:bootstrap} provides comparison of \SSH{} and A-SPAIN with corresponding convex approaches, showing the bootstrap 95\%
confidence intervals \cite{EfronTibshirani:Bootstrap} for the mean value of \snr{}.
It can be seen from the width of the confidence intervals that the comparison with Janssen would not be statistically significant, while, on the other hand, SPAIN clearly outperforms the $\ell_1$ methods in most cases.

\subsection{Evaluation based on PEMO-Q}

For further comparison of SPAIN with Janssen, we performed a second experiment, aiming at a~more profound evaluation using PEMO-Q \cite{Huber:2006a}.
In order to correctly evaluate the restored signals with PEMO-Q, only sound excerpts sampled at 44.1\,kHz from the previous experiment had to be used (which makes six test signals in total).
We also considered only the gaps with lengths from the set of $\{20,30,40,50\}$\,ms.
The measured quantity was the \emph{objective difference grade} (ODG) which simulates the human perception,
comparing the restored signal to the reference (original, not degraded signal).
The ODG ranges from $0$ to $-4$ and rates the audio degradation as:

\vspace{1ex}
\begin{tabular}{rl}
	 $0.0$ & Imperceptible \\
	$-1.0$ & Perceptible, but not annoying \\
	$-2.0$ & Slightly annoying \\
	$-3.0$ & Annoying \\
	$-4.0$ & Very annoying
\end{tabular} 
\vspace{1ex}

\begin{figure}[t]
	\begin{subfigure}{0.47\linewidth}
		\includegraphics[width=\textwidth]{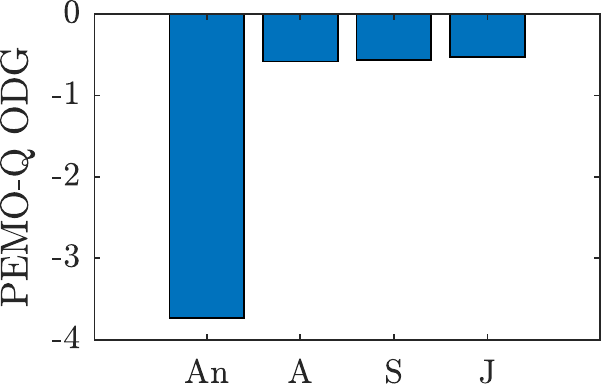}
		\caption{gap length 20\,ms}
	\end{subfigure}
    \hfill
	\begin{subfigure}{0.47\linewidth}
		\includegraphics[width=\textwidth]{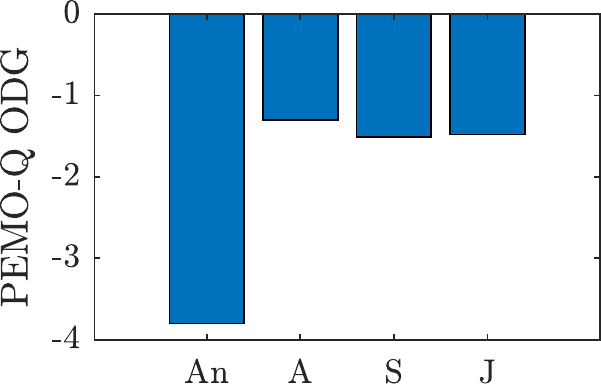}
		\caption{gap length 30\,ms}
	\end{subfigure}
	\\[1.1em]
	\begin{subfigure}{0.47\linewidth}
		\includegraphics[width=\textwidth]{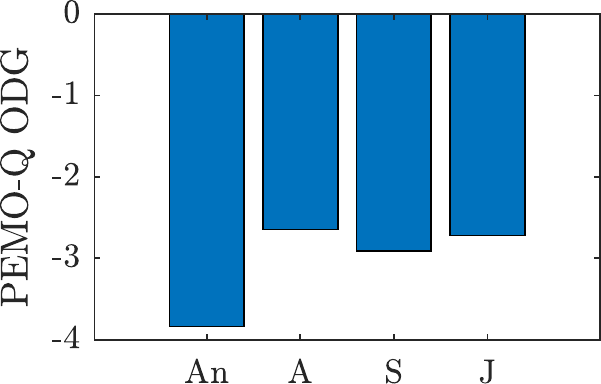}
		\caption{gap length 40\,ms}
	\end{subfigure}
    \hfill
	\begin{subfigure}{0.47\linewidth}
		\includegraphics[width=\textwidth]{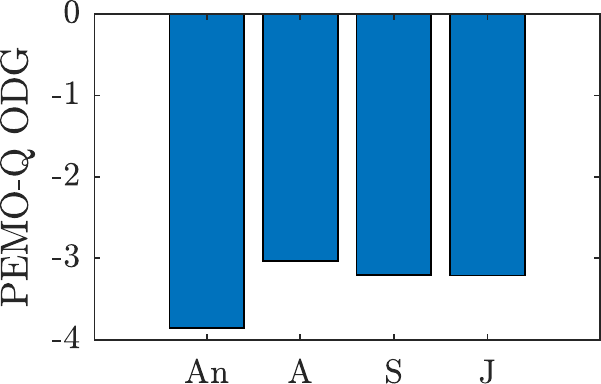}
		\caption{gap length 50\,ms}
	\end{subfigure}

	\caption{Average ODG values.
	Showing results for the observed, degraded signal (anchor, An) and the reconstruction by \mbox{A-SPAIN} (A), \SSH{} (S) and Janssen (J).}
	\label{fig:odg}
\end{figure}

The results in terms of ODG are presented in Fig.\,\ref{fig:odg}.
Each plot shows average values over the six audio samples, given the gap length.
As expected, the plots indicate that the longer the gap, the worse the reconstruction.
Nevertheless, the remarkable result here is that besides the gap length of 20\,ms, A-SPAIN slightly outperforms both \SSH{} and Janssen.



\section{Software and data}
The MATLAB codes for SPAIN and the sound excerpts are available at 
\url{https://bit.ly/2zdhbpp}.

\section{Conclusion}
\label{sec:conclusion}
The paper presented a novel inpainting algorithm (SPAIN) developed by an adaptation of successful declipping method, SPADE, to the context of inpainting.
It was shown that the analysis variant of SPAIN performs the best in terms of \snr{} among sparsity-based methods.
Furthermore, \mbox{A-SPAIN} was demonstrated to reach results on a par with the state-of-the-art Janssen algorithm for audio inpainting in terms of \snr{}.
Finally, the objective test using PEMO-Q, which takes into account the human perception of sound, showed that \mbox{A-SPAIN} even slightly outperforms Janssen for larger gap sizes.

\newpage


\newcommand{\noopsort}[1]{} \newcommand{\printfirst}[2]{#1}
\newcommand{\singleletter}[1]{#1} \newcommand{\switchargs}[2]{#2#1}

\end{document}